\documentclass[
a4paper,reprint,prb,
superscriptaddress,
amsmath,amssymb,
aps,twocolumns
]{revtex4-1}

\usepackage{graphicx}
\usepackage{dcolumn}
\usepackage{bm}
\usepackage{hyperref}
\usepackage{color}
\usepackage{dsfont}

\usepackage{color}

\allowdisplaybreaks[4]

\begin{document}

\title{Energy transport between critical one-dimensional systems\\ with different central charges}

\author{Sonja Fischer}
\affiliation{Institute for Theoretical Physics and Center for Extreme Matter and Emergent Phenomena, Utrecht University, Leuvenlaan 4, 3584 CE Utrecht, The Netherlands}
\author{Christoph Karrasch}
\affiliation{Technische Universit\"at Braunschweig, Institut f\"ur Mathematische Physik, Mendelssohnstraße 3, 38106 Braunschweig, Germany}
\author{Dirk Schuricht}
\affiliation{Institute for Theoretical Physics and Center for Extreme Matter and Emergent Phenomena, Utrecht University, Leuvenlaan 4, 3584 CE Utrecht, The Netherlands}
\author{Lars Fritz}
\affiliation{Institute for Theoretical Physics and Center for Extreme Matter and Emergent Phenomena, Utrecht University, Leuvenlaan 4, 3584 CE Utrecht, The Netherlands}

\begin{abstract}
Energy transport can reveal information about interacting many-body systems beyond other transport probes. In particular, in one dimension it has been shown that the energy current is directly proportional to the central charge, thus revealing information about the degrees of freedom of critical systems. In this work, we explicitly verify this result  in two cases for translationally invariant systems based on explicit microscopic calculations. More importantly, we generalise the result to non-translation invariant setups and use this to study a composite system of two subsystems possessing different central charges. We find a bottleneck effect meaning the smaller central charge limits the energy transport.
\end{abstract}

\maketitle

\section{Introduction}
Transport properties are one of the most fundamental ways to characterise and classify condensed matter systems.
Electrical transport properties, for instance, can distinguish metals from insulators or superconductors.
Very specific electrical transport properties can further fan out the classification, for example, insulating bulk systems can have interesting boundary transport properties, as is the case in quantum Hall systems, topological insulators, or other topologically ordered systems.

Although generally more complicated to measure and compute, the same is true for heat transport. Additional difficulties arise in this case from the fact that all degrees of freedom participate in the transport process and not only those associated with a charge under the electromagnetic field, ie, phonons and collective modes also contribute. In the early days of research on metals it was established that to an excellent degree of approximation those systems fulfilled the so-called Wiedemann--Franz law.\cite{WiedemannFranz53} This law states that as the temperature $T$ goes to zero, the ratio of the electrical conductivity $\sigma$ and the heat conductivity $\kappa$ tends to $\sigma/\kappa = T L_0$ where $L_0=\pi^2k_\text{B}^2/(3e^2)$ is the Lorenz number, a constant only composed out of fundamental quantities. Nowadays, the Wiedemann--Franz law or the violation thereof\cite{Wakeham-11,Crossno-16,Dutta-17} serves as a diagnostic tool for the existence or non-existence of a Fermi liquid state.

In one-dimensional systems, more recent work\cite{BernardDoyon12} uncovered a fundamental connection between the central charge of a critical system with conformal symmetry and its heat-transport properties. If the system is characterised by a central charge $c$, but it consists of two subsystems, left and right, held at temperatures $T_\text{L}$ and $T_\text{R}$ respectively, the energy current is given by
\begin{eqnarray}\label{eq:conjecture}
J_\text{E}=\frac{c\pi}{12} \left(T_\text{L}^2-T_\text{R}^2 \right),
\end{eqnarray}
and thus directly proportional to the central charge itself. 
The central charge can be interpreted loosely as a measure for the number of degrees of freedom in a theory. It describes how a system behaves in the presence of a macroscopic length scale, for example the Casimir energy is directly proportional to $c$. At a more formal level, the central charge dictates the short-distance behaviour of the correlation functions of the energy-momentum tensor and appears as a quantum anomaly.\cite{DiFrancescoMathieuSenechal97,Mussardo10} It is important to note that for the result \eqref{eq:conjecture} to hold, overall translational invariance of the equilibrium situation $T_\text{L}=T_\text{R}$ is required. The result \eqref{eq:conjecture} was subsequently confirmed in free systems like the transverse-field Ising chain\cite{DeLuca-13,ColluraKarevski14,Kormos17,PerfettoGambassi17} as well as interacting systems using the generalised hydrodynamic approach\cite{Castro-Alvaredo-16,Bertini-16,BertiniPiroli18,Horvath19} and numerical simulations.\cite{Karrasch-13,DeLuca-14,Biella-16}

The purpose of this paper is two-fold: we first verify the result \eqref{eq:conjecture} for two microscopic models corresponding to field theories with $c=1$ and $c=1/2$, ie, free fermions and free Majorana fermions, respectively.
Additionally, we study a situation in which we couple two subsystems with these two different central charges.
As our main result we find a bottleneck on the energy current, namely the subsystem with the smaller central charge limits the energy transport.
The convenient framework to study this is a one-dimensional version of a spinless p-wave superconductor, which can realise central charges of $c=1$ and $c=1/2$ at its respective XX and Ising critical point.
We will use two complementary methods: (i) the non-equilibrium Green function technique,\cite{RammerSmith86,Kamenev11} which allows us to obtain exact analytical results in the thermodynamic limit at all parameter regimes, and (ii) the real-time\cite{Vidal04,WhiteFeiguin04,Daley-04,Schmitteckert04} finite-temperature\cite{FeiguinWhite05,Barthel-09} density matrix renormalisation group (DMRG)\cite{White92,Schollwoeck11} algorithm\cite{Karrasch-12,Karrasch-13NJP,Karrasch-13} which works directly with the corresponding lattice models.

The organisation of the paper is as follows: we first introduce the model in Sec.~\ref{sec:model} and briefly review its key properties. Sec.~\ref{sec:energycurrent} discusses the basic the energy current operator in its most generic form and relates it to Green functions, while Sec.~\ref{sec:dmrg} gives details on the DMRG simulations. In Sec.~\ref{sec:main} we present our main results and finish with a conclusion in Sec.~\ref{sec:conclusion}. Technical details are deferred to the appendices.

\section{The Model}\label{sec:model}
\begin{figure}[t]
	\centering
	\includegraphics[width=0.7\columnwidth]{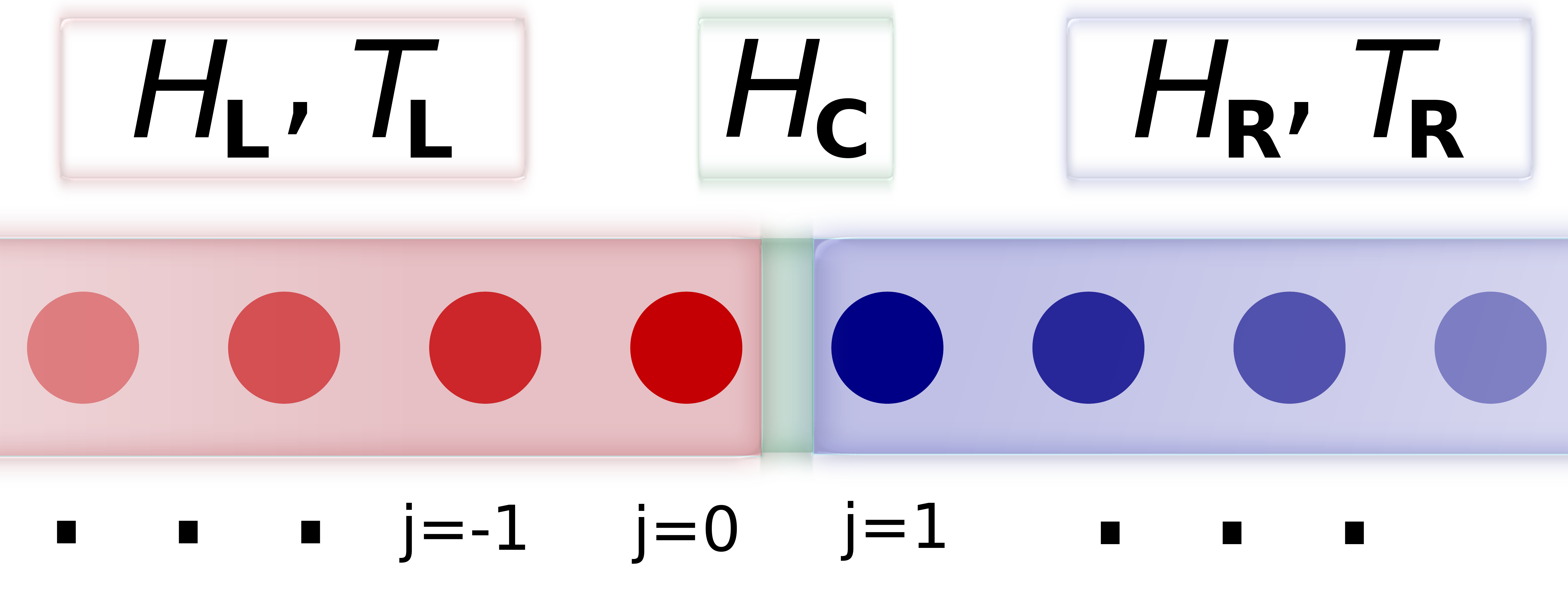}
	\caption{Setup considered in this work: Two semi-infinite fermionic chains described by $H_\text{L}$ and $H_\text{R}$ at different temperatures $T_\text{L}$ and $T_\text{R}$ are joined at their endpoints with a coupling Hamiltonian $H_\text{C}$. The Hamiltonians $H_\text{L/R}$ may contain p-wave superconducting pairing terms.}
    \label{fig:setup}
\end{figure}
The setup we consider consists of two subsystems, henceforth referred to as `left' (L) and `right' (R). Both are held at different but constant temperatures $T_\text{L}$ and $T_\text{R}$ and are described by Hamiltonians $H_\text{L}$ and $H_\text{R}$, respectively. The two subsystems are coupled by a Hamiltonian $H_\text{C}$ which facilitates energy  flow between them, see Fig.~\ref{fig:setup} for a sketch of the setup. Specifically we consider a Hamiltonian of the type
\begin{eqnarray}
H&=& H_\text{L}+H_\text{R}+H_\text{C},\label{eq:ham}\\
H_\text{L}&=&
-t_\text{L} \sum_{j=-N_\text{L}}^{-1} \left( c_j^\dagger c_{j+1}^{\phantom{\dagger}}+c_{j+1}^\dagger c_{j}^{\phantom{\dagger}}\right)
-\epsilon_\text{L} \sum_{j=-N_\text{L}}^0 c^\dagger_j c^{\phantom{\dagger}}_j \nonumber\\*
& &-\Delta_\text{L} \sum_{j=-N_\text{L}}^{-1} \left(c_{j}^\dagger c_{j+1}^\dagger+c_{j+1}^{\phantom{\dagger}}c_j^{\phantom{\dagger}}\right),\\
H_\text{R}&=&-t_\text{R} \sum_{j=1}^{N_\text{R}-1} \left( c_j^\dagger c_{j+1}^{\phantom{\dagger}}+c_{j+1}^\dagger c_{j}^{\phantom{\dagger}}\right) -\epsilon_\text{R} \sum_{j=1}^{N_\text{R}} c^\dagger_j c^{\phantom{\dagger}}_j\nonumber\\*
& &-\Delta_\text{R} \sum_{j=1}^{N_\text{R}-1} \left(c_{j}^\dagger c_{j+1}^\dagger+c_{j+1}^{\phantom{\dagger}}c_j^{\phantom{\dagger}}\right), \\
H_\text{C}&=&-t_\text{C}  \left( c_0^\dagger c_{1}^{\phantom{\dagger}}+c_{1}^\dagger c_{0}^{\phantom{\dagger}}\right) -\Delta_\text{C}  \left(c_{0}^\dagger c_{1}^\dagger+c_{1}^{\phantom{\dagger}}c_0^{\phantom{\dagger}}\right)\label{eq:HamC}.
\end{eqnarray}
This Hamiltonian describes a one-dimensional p-wave superconductor with hopping parameters $t_\text{L},t_\text{R},t_\text{C}$ and pairing terms $\Delta_\text{L},\Delta_\text{R},\Delta_\text{C}$ in their respective parts of the chain. The operator $c_j$ is an annihilation operator for a spinless fermion at site $j$ where negative (positive) $j$ indicate a site in the left (right) subsystem (site $0$ belongs to the left subsystem). We note that the total number of lattice sites equals $N_\text{L}+N_\text{R}+1$. Additionally, we introduce on-site energies $\epsilon_\text{L}$ and $\epsilon_\text{R}$.

The main motivation to study this Hamiltonian is that it allows for a variety of different situations including that of having different central charges $c_\text{L,R}=1/2,1$ in the respective subsystems, as well as different speeds of light. For example, the $c_\text{L}=1/2$ critical point is obtained for $t_\text{L}=\Delta_\text{L}=\epsilon_\text{L}/2$, with the velocity of the low-lying excitations given by $v=2t_\text{L}$. On the other hand, $c_\text{L}=1$ is obtained for the choice $\Delta_\text{L}=\epsilon_\text{L}=0$ with the velocity $v=2t_\text{L}$. For most other parameter sets this Hamiltonian describes a gapped system which one can also study, as done in Sec.~\ref{sec:Ising}. These situations, however, are of little interest to our discussion since at this point a comparison to the result Eq.~\eqref{eq:conjecture} is impossible. If one is looking into experimental situations, this type of Hamiltonian can be encountered in a variety of settings. Apart from direct electronic implementations it can also appear as an effective model of spin chains via a Jordan--Wigner transformation\cite{Giamarchi04} (see Appendix~\ref{app:spinchains}).

\section{Green function formalism}\label{sec:energycurrent}
The heat current is the temporal change of heat in either of the two subsystems. It consists of two contributions, namely a change in the energy and one related to the particle flow. Choosing the left lead as reference,  it is given by\cite{Giazotto-06} $Q_\text{L}=J_\text{E}-\mu \dot{N}_\text{L}$ where $Q_\text{L}$ is the heat flowing out of the left lead, while $J_\text{E}=-\dot{E}_\text{L}$ and $-\dot{N}_\text{L}$ are the energy and particle flows, respectively, and $\mu$ denotes an external chemical potential. In the remainder of this work we only consider the case $\mu=0$ implying the heat current is equal to the temporal change of energy in the lead, ie, $Q_\text{L}=J_\text{E}=-\dot{E}_\text{L}$. Furthermore, we are only interested in the steady state heat current where we have $Q_\text{L}=-Q_\text{R}$.

In order to derive the energy current in the left lead we use the Heisenberg equations of motion following $ \dot{E}_\text{L}=\frac{\text{d}}{\text{d}t}\langle H_\text{L}\rangle=\text{i}\langle[H,H_\text{L}]\rangle$ (note that $\hbar=1$ throughout). Straightforward computations allow to rewrite this in terms of the fermionic operators as
\begin{eqnarray}
        \dot{E}_\text{L}&=&-\text{i}\left(t_\text{L} t_\text{C}-\Delta_\text{L} \Delta_\text{C} \right)\left\langle c^{\dagger}_{-1}c^{\phantom{\dagger}}_1+c^{\phantom{\dagger}}_{-1} c^{\dagger}_{1}\right\rangle\nonumber\\
            & &-\text{i} \left(t_\text{L} \Delta_\text{C}-t_\text{C} \Delta_\text{L} \right) \left\langle c_{-1}^{\phantom{\dagger}}c_1^{\phantom{\dagger}}+c_{-1}^\dagger c_{1}^\dagger \right\rangle\nonumber \\
            & &-\text{i} t_\text{C} \epsilon_\text{L} \left\langle c_{0}^{\phantom{\dagger}}c_1^\dagger +c_{0}^{\dagger}c_1^{\phantom{\dagger}} \right\rangle-\text{i} \Delta_\text{C} \epsilon_\text{L} \left\langle c_{0}^{\phantom{\dagger}} c_1^{\phantom{\dagger}}+ c_0^{\dagger}c_{1}^\dagger\right\rangle.\quad\label{eq:current0}
\end{eqnarray}
It is important to note here that apart from `normal' fermionic Green functions this also requires knowledge of anomalous propagators due to the superconducting correlations. The calculations are carried out in a non-equilibrium setting and we consequently resort to the Schwinger--Keldysh closed time contour formalism,\cite{Kamenev11} which we use to derive exact results in the thermodynamic limit. We note that this requires the system to be an effectively free theory, in our case corresponding to $c=1/2$ and $c=1$ respectively. In contrast, the application of the non-equilibrium Green function technique to interacting theories would require approximations. Using the results and definitions from Appendix~\ref{sec:genfun} [especially Eq.~\eqref{eq:expectationvalues1}--\eqref{eq:expectationvalues4}] the above expectation values can be expressed in terms of lesser Green functions according to
\begin{eqnarray}
        \dot{E}_\text{L} & = &\phantom{+}
        \left(t_\text{L} t _\text{C}- \Delta_\text{L} \Delta_\text{C} \right)\text{Re}\left[ G^{<}_{1,-1}(t,t) + \bar{G}^<_{1,-1}(t,t) \right]\nonumber\\
                  & & + \left(t_\text{L} \Delta_\text{C} - \Delta_\text{L} t_\text{C} \right) \text{Re} \left[ F^{<}_{1,-1}(t,t) + \bar{F}^{<}_{1,-1}(t,t) \right]\nonumber\\
                  & & + t_\text{C} \epsilon_\text{L} \text{Re}\left[ G^{<}_{1,0}(t,t)+ \bar{G}^{<}_{1,0}(t,t)\right]\nonumber\\
                  & & + \Delta_\text{C} \epsilon_\text{L} \text{Re} \left[F^{<}_{1,0}(t,t)+\bar{F}^{<}_{1,0}(t,t)  \right],
\end{eqnarray}
where $G^<$ and $\bar{G}^<$ denote the normal particle and hole propagators and $F^<$ and $\bar{F}^<$ describe anomalous propagators for Cooper pairs and anti-Cooper pairs respectively [see Eq.~\eqref{eq:nambustructure} below]. Restricting ourselves to a steady state situation allows to apply Fourier transformation from time into frequency domain. Furthermore, we can significantly simplify the expression for the current using the Dyson equation for the lesser Green functions.

\begin{figure}[t]
	\centering
	\includegraphics[width=0.6\columnwidth]{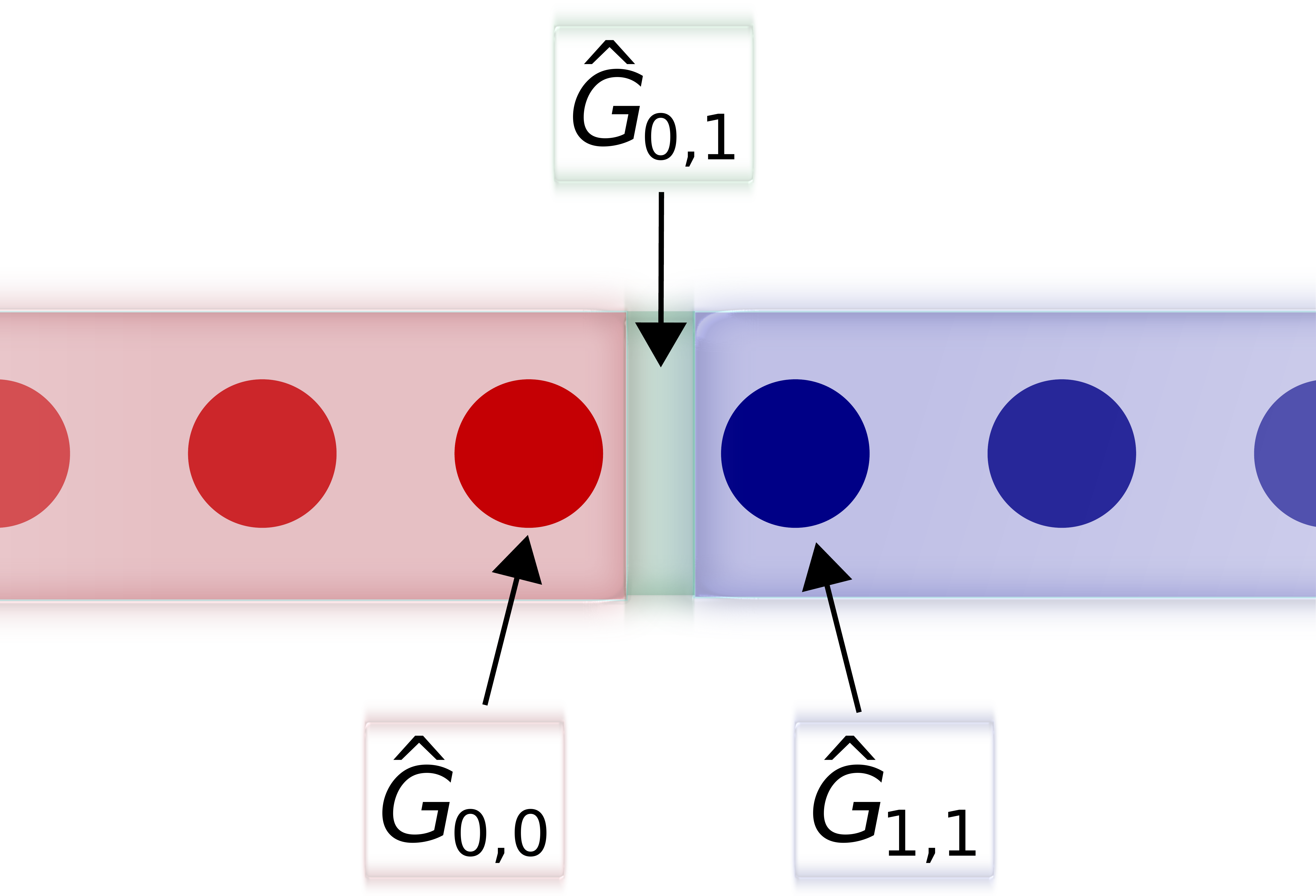}
	\caption{Graphical representation of the Green functions relevant for transport, see Eq.~\eqref{eq:G10}. The Green function $\hat{G}_{1,0}$ can be expressed by the Green functions $\hat{G}_{0,0}$ and $\hat{G}_{1,1}$}
	\label{fig:setup2}
\end{figure}
Since lesser Green functions are off-diagonal in terms of the time contour description, the Dyson equation for the lesser Green function reads
\begin{equation}
\label{eq:dyson}
    \sum_{j}\hat{G}^{<}_{i,j} \left(\omega\,\delta_{j,k}-\mathcal{H}_{j,k}\right)=0,
\end{equation}
where the Green function $\hat{G}$ is a matrix in Nambu space that has the structure
\begin{equation}\label{eq:nambustructure}
    \hat{G}_{j,k}(\omega)=\begin{pmatrix} G & F \\   \bar{F} & \bar{G} \end{pmatrix}_{j,k}(\omega)
\end{equation}
and $\mathcal{H}_{j,k}$ denotes the matrix representation of the Hamiltonian; see Eq.~\eqref{eq:hamextended} for the precise form. Using Eq.~\eqref{eq:dyson} in the case of $i=1$ and $k=0$, we can rewrite the energy current in a compact form as
\begin{equation}\label{eq:curr10}
    \begin{split}
        \dot{E}_\text{L} = & -\int \frac{\text{d}\omega}{2 \pi} \text{Re}\left\{\omega \left[t_\text{C}  \left(G -\bar{G}\right)_{1,0} - \Delta_\text{C} \left(F-\bar{F}\right)_{1,0}\right]^{<}\right.\\
                                               & \left.\!\!\!\!+\left[ \left(t_\text{C}^2-\Delta_\text{C}^2\right)\left(G+\bar{G}\right)_{1,1} + 2t_\text{C}\Delta_\text{C}\left( F+\bar{F}\right)_{1,1}\right]^{<}\right\}.
    \end{split}
\end{equation}
Unless specified the integration ranges over the real axis, $-\infty<\omega<\infty$. Also from now on the Green functions are understood to depend on the frequency $\omega$. The second line of Eq.~\eqref{eq:curr10} does not contribute since the local lesser Green functions do not have a real part and thus we can drop them. Importantly, this expression depends solely on the Green function that describes the transition from the last side of the left lead to the first side of the right lead, whereas the rest of the system does not contribute explicitly. Furthermore, using the results of Appendix~\ref{sec:todo}, we can decompose this element of the Green function into two parts that only contain local Green functions situated at the ends of each lead according to
\begin{equation}
\label{eq:G10}
    \hat{G}_{1,0} = -\hat{G}^{(0)}_{1,1}\,\hat{t}_\text{C}\,\hat{G}_{0,0},
\end{equation}
which is graphically represented in Fig.~\ref{fig:setup2}. Specifically, the transition Green function consists of one bare Green function $\hat{G}^{(0)}_{1,1}$, one full Green function $\hat{G}_{0,0}$, and a coupling matrix $\hat{t}_\text{C}$ that is given by
\begin{equation}
\hat{t}_\text{C} = \begin{pmatrix}
t_\text{C}& \Delta_\text{C} \\ -\Delta_\text{C} & -t_\text{C}
\end{pmatrix}.
\end{equation}
In this context the bare Green function $\hat{G}^{(0)}_{1,1}$ is the Green function of the first site of the right chain with the left chain decoupled, whereas the full Green function $\hat{G}_{0,0}$ is the one of the last site of the left lead computed in the presence of a coupling to the right lead. It is worthwhile noting that the choice of 'bare' for the right lead and 'full' for the left lead is arbitrary and could be switched around.

Although it is possible to perform all the calculations for the full model \eqref{eq:ham}, for the sake of brevity we specify below to the relevant cases possessing conformal symmetry.

\section{DMRG formalism}\label{sec:dmrg}
In order to compute the energy current $J_\textnormal{E}$ using the DMRG method,\cite{White92,Schollwoeck11} we consider the setup shown in Fig.~\ref{fig:setup} as an initial condition and subsequently perform a time-evolution until a stationary state is reached. In practice, one can reach only finite time scales (see below), which are typically of the order of several $\mathcal{O}(10/t_{\textnormal{L,R}})$. This leads to a finite-time error whose magnitude one can estimate, eg, by interchanging $T_\textnormal{L}\leftrightarrow T_\textnormal{R}$ and comparing results (which agree only in the limit $t\to\infty$).

To be more precise, the thermal density matrices $e^{-H_{\textnormal{L,R}}/T}$ as well as the real-time evolution operators $e^{-iHt}$ are determined using a time-dependent DMRG algorithm; both operators are factorized by a fourth order Trotter-Suzuki decomposition.\cite{Vidal04,WhiteFeiguin04,Daley-04,Schmitteckert04} We incorporate finite temperatures using the purification technique.\cite{FeiguinWhite05,Barthel-09} The discarded weight during each individual `bond update' is kept below a pre-defined discarded weight, which leads to an exponential increase of the bond dimension during the real-time evolution. In order to access time scales as large as possible, we employ a finite-temperature disentangler, \cite{Karrasch-12,Karrasch-13NJP} which exploits the fact that purification is not unique to slow down the growth of the bond dimension. Our calculations are performed using a system size of the order of $N_\textnormal{L}=N_\textnormal{R}=128$ sites. The Hamiltonian is transformed into a spin representation via a Jordan-Wigner transformation (see Appendix \ref{app:spinchains}).

\section{Main results for the energy current}\label{sec:main}
In the following we discuss three different cases:  (a) Coupling two subsystems with effective Majorana degrees of freedom, ie, $c=1/2$ in both subsystems. (b) Coupling two free fermion systems, ie, $c=1$ in both subsystems. (c) Coupling a Majorana fermion to a fermion system, meaning $c=1/2$ and $c=1$ in the respective subsystems. We note that while the first two cases realise a setup covered by Eq.~\eqref{eq:conjecture} at their respective massless points, provided there is overall translational symmetry of the parameters of the Hamiltonian, in the latter situation the result Eq.~\eqref{eq:conjecture} is not applicable. Furthermore, we simplify the setup by always choosing the parameters of the coupling region, described by Eq.~\eqref{eq:HamC}, to equal the parameters of the right lead, ie, $t_\text{C}=t_\text{R}$ and $\Delta_\text{C}=\Delta_\text{R}$.

\subsection{Coupling two Majorana chains}\label{sec:Ising}
In this section we consider the two Majorana fermion systems. This is obtained by setting $\Delta_{a}=t_{a}$, $a=\text{L/R}$. However, we allow for the parameters on the left and right side to be different, ie, $t_\text{L}\neq t_\text{R}$, thus breaking translational invariance. Furthermore, for $\epsilon_a=2t_a$ the system is gapless with the velocity of the low-lying excitations given by $v=2t_a$ and the central charge $c_a=1/2$. On the other hand, for $\epsilon_a\neq 2t_a$ the system possesses an energy gap. We first calculate the energy current for general $\epsilon_a$ and specify to the gapless case in the end.

Using Eq.~\eqref{eq:G10} we can simplify the expression for the current, Eq.~\eqref{eq:curr10}, to
\begin{eqnarray}
    \dot{E}_\text{L}&=& -\frac{t^2_\text{R}}{2\pi}\int \text{d} \omega\,\omega\,\text{Re}\left(\mathcal{R}^{(0)}\mathcal{L}\right)^<\\
            &=& - \frac{t^2_\text{R}}{2\pi}\int \text{d} \omega\,\omega\,\text{Re}\left(\mathcal{R}^{(0)R}\mathcal{L}^{<}+\mathcal{R}^{(0)<}\mathcal{L}^{A}\right),\label{eq:Isingcurrent}
\end{eqnarray}
where we defined $\mathcal{R}^{(0)}=(G+\bar{G}+F+\bar{F})^{(0)}_{1,1}$ and $\mathcal{L}=(G+\bar{G}-F-\bar{F})_{0,0}$ and applied the Langreth theorem.\cite{HaugJauho08} The combination of Green functions $\mathcal{R}^{(0)}$ are composed of bare Green functions of the right lead, whereas, as has been stated in the previous section, the combination $\mathcal{L}$ is made up of full Green functions of the left lead, ie, it contains information about the coupling.

In the following we adopt the convention that if not explicitly indicated by a superscript, the Green function is the retarded component. For a semi-infinite lead we can derive a self-consistency equation, as shown in Appendix C, see Eq.~\eqref{eq:selfconsistencyr}, which reduces to
\begin{equation}
    \hat{G}^{(0)}_{1,1}=\left[\left(\hat{G}^{(0)-1}\right)_{1,1} - t_\text{R}^2 \mathcal{R}^{(0)}\begin{pmatrix} 1& -1 \\ -1 & 1 \end{pmatrix}\right]^{-1}.
\end{equation}
Adding up all matrix elements of $G_{1,1}^{(0)}$ according to the definition of $\mathcal{R}^{(0)}$ and solving the resulting quadratic equation for $\mathcal{R}^{(0)}$ yields
\begin{equation}\label{eq:rdef}
    \mathcal{R}^{(0)} =\frac{1}{|t_\text{R}|}\left(\omega_\text{R} -\text{i} \rho_\text{R}\right),
\end{equation}
where we have introduced the dimensionless energy scale $\omega_\text{R}=(\omega^2+4t_\text{R}^2- \epsilon_\text{R}^2)/(4\omega |t_\text{R}|)$ as well as the dimensionless density of states $\rho_\text{R}={\rm{Re}}\sqrt{1-\omega_\text{R}^2} +\pi \delta(\omega) \frac{4t_\text{R}^2-\epsilon_\text{R}^2}{4|t_\text{R}|}\theta(4t_\text{R}^2/\epsilon_\text{R}^2-1)$.

At $\omega=0$, the density of states $\rho_\text{R}$ has a pole manifesting the Majorana edge mode present in the topological phase for $|\epsilon_\text{R}|<2t_\text{R}$. However, in the current Eq.~\eqref{eq:Isingcurrent}, we need to consider the combination $\omega\,\text{Re}(\mathcal{R}^{(0)} \mathcal{L}^{<}+\mathcal{R}^{(0)<}\mathcal{L}^{*})$ with $^*$ denoting the usual complex conjugation. Using that even though $\mathcal{R}^{(0)}$ is divergent, $\omega \mathcal{R}^{(0)}$ and $\omega \mathcal{L}^{(0)}$ are perfectly well defined for all values of $\omega$, we can approximate $\mathcal{N}$ in Eq.~\eqref{eq:lret} as
\begin{equation}
    \mathcal{N}(\omega=0)\approx 4t_\text{L}^2t_\text{R}^2 \mathcal{L}^{(0)}\mathcal{R}^{(0)}
\end{equation}
and therefore find
\begin{eqnarray}
    \left|\omega \mathcal{R}^{(0)<}\mathcal{L}^{*}\right|_{\omega\rightarrow 0} &\approx& \left|\frac{\omega(-2\text{i} f_\text{R} \text{Im} \mathcal{R}^{(0)})(-4t_\text{L}^2\mathcal{L}^{(0)})}{4t_\text{L}^2t_\text{R}^2 \mathcal{R}^{(0)}\mathcal{L}^{(0)}}\right|_{\omega=0}\nonumber\\
    &=&0.
\end{eqnarray}
By the same logic, the technically more involved term $\omega \mathcal{L}^< \mathcal{R}^{(0)}$ yields the same vanishing result. For the energy current the precise properties of the leads at zero energy are therefore irrelevant since these modes do not carry any energy.  Therefore we can safely ignore the  $\delta$-function in Eq.~\eqref{eq:rdef} and use $\rho_\text{R}={\rm{Re}}\sqrt{1-\omega^2_\text{R}}$.

The solution to the left-side semi-infinite lead can be obtained in a similar fashion using Eq.~\eqref{eq:selfconsistencyl} and is given by $\mathcal{L}^{(0)}=\frac{1}{|t_\text{L}|}\left(\omega_\text{L}-\text{i}\rho_\text{L}\right)$ with $\omega_\text{L}=(\omega^2+4t_\text{L}^2-\epsilon_\text{L}^2)/(4\omega|t_\text{L}|)$ and $\rho_\text{L}={\rm{Re}}\sqrt{1-\omega_\text{L}^2}$. In the latter expression we have already neglected the possible Majorana quasiparticle pole. Since the uncoupled leads are kept at a fixed temperature, one can simply extract the lesser Green functions of a bare lead via $\mathcal{R}^{(0)<}=\text{i} f_\text{R} \left(\mathcal{R}^{(0)R}-\mathcal{R}^{(0)A}\right)$ and $\mathcal{L}^{(0)<}=\text{i}f_\text{L}(\mathcal{L}^{(0)R}-\mathcal{L}^{(0)A})$ where the superscripts $R$ and $A$ denote the retarded and advanced Green functions and $f_\text{L/R}(\omega)=\left[1+\exp \left(\omega/T_\text{L/R}\right)\right]^{-1}$ are the Fermi distribution functions corresponding to the respective temperatures of each lead.

In the equation for the energy current \eqref{eq:Isingcurrent} we also need the full Green function of the coupled left lead. Using Eq.~\eqref{eq:fullgf0} for $i=0$ together with the results for $\mathcal{L}^{(0)}$ and $\mathcal{R}^{(0)}$ yields
\begin{equation}
    \begin{split}\label{eq:lret}
        \mathcal{L}&=\frac{2\omega-4t_\text{L}^2\mathcal{L}^{(0)}}{\mathcal{N}},\\
        \mathcal{N}&=\omega^2-\epsilon_\text{L}^2 - 2 \omega \left(t_\text{L}^2\mathcal{L}^{(0)}+t_\text{R}^2\mathcal{R}^{(0)}\right) + 4t_\text{L}^2t_\text{R}^2\mathcal{L}^{(0)}\mathcal{R}^{(0)}.
    \end{split}
\end{equation}
Since the system is not in thermal equilibrium, finding the lesser Green function for the full left lead proves to be more difficult than for the bare system. However, in the steady state, we can calculate it via
\begin{equation}\label{eq:gldef}
\hat{G}_{0,0}^<=\hat{G}^R_{0,0}\hat{\Sigma}^{<}\hat{G}^{A}_{0,0}
\end{equation}
 with the lesser self energy given by
\begin{equation}\label{eq:lesserse}
\hat{\Sigma}^<= t_\text{R}^2 \mathcal{R}^{(0)<}\begin{pmatrix} \phantom{-}1&-1 \\ -1 &\phantom{-} 1 \end{pmatrix} + t^2_\text{L} \mathcal{L}^{(0)<}\begin{pmatrix}\phantom{-}1&\phantom{-}1\\\phantom{-} 1&\phantom{-}1 \end{pmatrix}.
\end{equation}
Using the fact that taking the real part in Eq.~\eqref{eq:Isingcurrent} restricts us to a regime where $\omega_\text{L/R}^2<1$, ie, $\rho_\text{L/R}^2=1-\omega_\text{L/R}^2$, a tedious but straightforward calculation allows us to write the final expression for the energy current as
\begin{equation}
 \dot{E}_\text{L}=\frac{1}{\pi}\int\frac{ \text{d} \omega\,\omega\,t_\text{L} t_\text{R} \rho_\text{L} \rho_\text{R} \left[f_R(\omega)-f_L(\omega)\right]}{\left(t_\text{L}\omega_\text{L}-t_\text{R} \omega_\text{R}\right)^2 + \left(t_\text{L} \rho_\text{L} + t_\text{R} \rho_\text{R}\right)^2}.\label{eq:Isingcurrent2}
\end{equation}
We note that the densities $\rho_\text{L/R}$ depend on the frequency $\omega$ and in this way also restrict the range of integration. The result Eq.~\eqref{eq:Isingcurrent2} has the structure of a heat current derived by the Landauer--B\"uttiker formalism with the effective density of states $\rho_{\text{eff}}=(t_\text{L} t_\text{R} \rho_\text{L} \rho_\text{R})/[\left(t_\text{L}\omega_\text{L}-t_\text{R} \omega_\text{R}\right)^2 + \left(t_\text{L} \rho_\text{L} + t_\text{R} \rho_\text{R}\right)^2]$. In general the temperature dependence is governed by the existence or absence of an energy gap. In the former case one finds exponentially suppressed energy currents, while the latter case will be discussed in detail in the following.

Now let us specialise the general result \eqref{eq:Isingcurrent2} to the case of two gapless Majorana fermion systems. This is done by setting $\epsilon_\text{L/R}=2t_\text{L/R}$. Additionally taking the low-temperature limit $t_\text{L/R}/T_\text{L/R}\rightarrow \infty$, the energy current Eq.~\eqref{eq:Isingcurrent2} can be evaluated using a standard Sommerfeld expansion with the result
\begin{equation}
 \dot{E}_\text{L}=\frac{\pi}{6}\frac{t_\text{L}/t_\text{R}}{(1+t_\text{L}/t_\text{R})^2}\left(T_\text{R}^2-T_\text{L}^2\right).
 \label{eq:Isingfinal}
\end{equation}
Thus the energy current at low temperatures becomes maximal in the translationally invariant case $t_\text{L}=t_\text{R}$. In this case it also equals the result from conformal field theory \eqref{eq:conjecture} (recall $J_\text{E}=-\dot{E}_\text{L}$) since the central charge of this is given by $c=1/2$.

\begin{figure}[t]
\begin{center}
\includegraphics[width=0.45\textwidth]{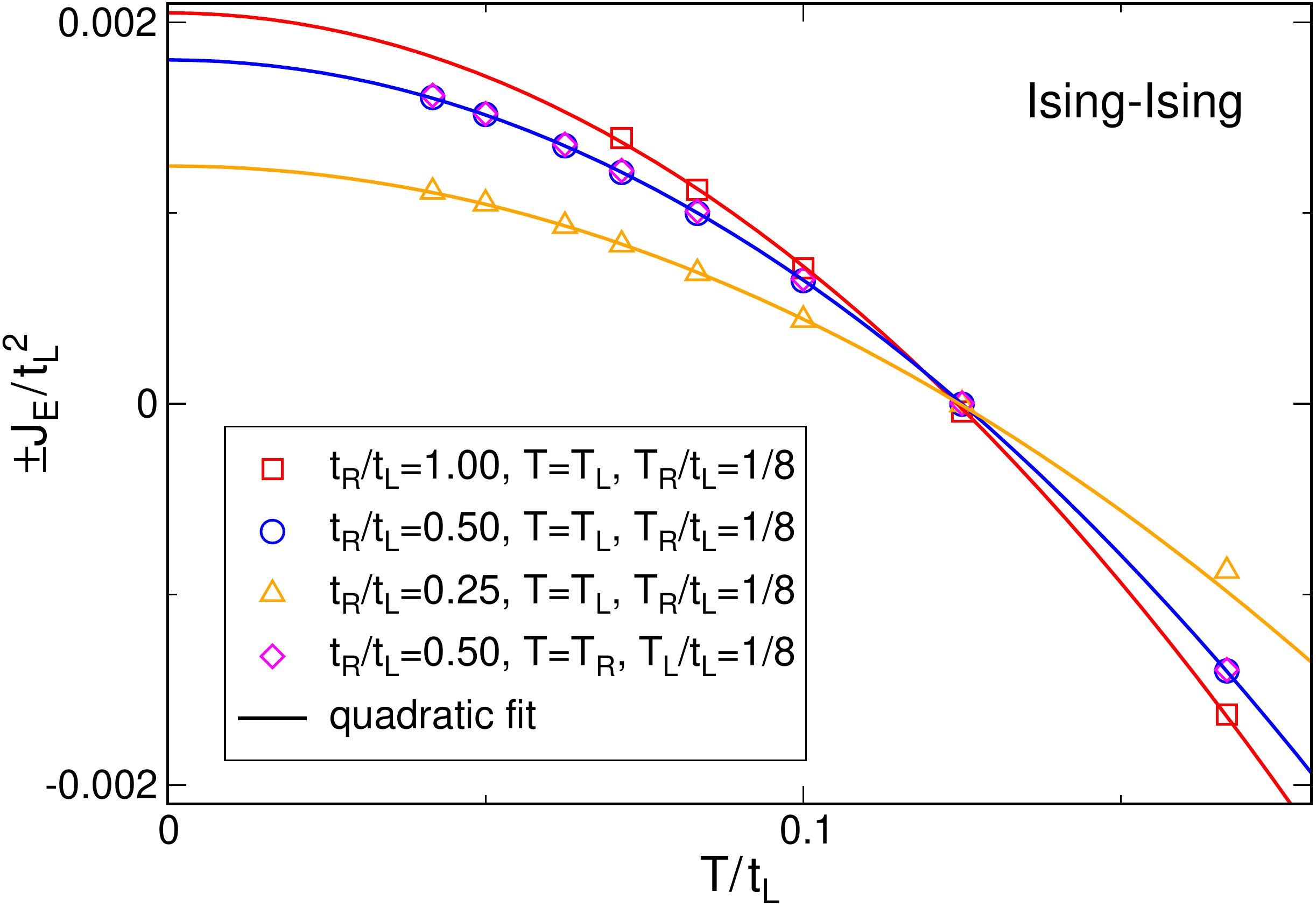}
\caption{Steady-state energy current for the coupling of two critical Ising chains with parameters $t_\text{L/R}$ as a function of the temperature in the left or right subsystem. The prefactor in \eqref{eq:Isingfinal} is obtained from a quadratic fit, the obtained values are collected in Table~\ref{tab:IsingIsing}.}
\label{fig:IsingIsing}
\end{center}
\end{figure}

The result \eqref{eq:Isingfinal} can be compared to DMRG simulations of the energy current. The energy currents obtained in the steady state are shown in Fig.~\ref{fig:IsingIsing} (as discussed in Sec.~\ref{sec:dmrg}, the finite-time error can be estimated by interchanging the temperatures $T_\textnormal{L}\leftrightarrow T_\textnormal{R}$ and comparing the results). We extract the prefactors from a quadratic fit (see Table~\ref{tab:IsingIsing}); they are in excellent agreement with the analytic prediction \eqref{eq:Isingfinal}.

\begin{table}[b]
\caption{Comparison of the analytic and numerical results for the prefactor of the quadratic temperature dependence of the energy current \eqref{eq:Isingfinal}.}
\begin{center}
\begin{tabular}{|c|c|c|}
\hline
$t_\text{L}/t_\text{R}$ & analytic value & fitted value\\
\hline
1 & 0.131 & 0.133 \\
2 & 0.116 & 0.116 \\
4 & 0.084 & 0.080 \\
\hline
\end{tabular}
\end{center}
\label{tab:IsingIsing}
\end{table}

\subsection{Coupling two fermionic chains}
A second case worthwhile investigating is the limit of vanishing pairing terms everywhere, ie, $\Delta_{a}=0$ with $a=\text{L/R}$. Here, our system reduces to a simple tight-binding chain of spinless fermions. In this regime all off-diagonal terms in the Nambu structure \eqref{eq:nambustructure} of all Green functions vanish. In principle it is also possible to reduce the $(2\times 2)$-matrix structure that still describes uncoupled particles and holes into a scalar structure that only encompasses particles. However, for the sake of later compatibility it is useful to keep the Nambu structure intact. Following the same logic as in the previous case, we first rewrite the current in terms of local Green functions according to
\begin{equation}\label{eq:xxcurrent}
    \dot{E}_\text{L}=- \frac{t^2_\text{R}}{2\pi}\int \text{d} \omega\,\omega\,\text{Re}\left(G^{(0)}_{1,1} G_{0,0} + \bar{G}_{1,1}^{(0)}\bar{G}_{0,0}\right)^<.
\end{equation}
In this case, the current does not depend on the combination of Green functions on the left and right side, $\mathcal{R}$ and $\mathcal{L}$, but it can be solved in the particle and hole channel separately. Using the self-consistency condition Eq.~\eqref{eq:selfconsistencyr} for the bare Green function of the right lead we find
\begin{eqnarray}
    G^{(0)}_{1,1}&=&\frac{1}{|t_\text{R}|}\left(\omega^+_\text{R}-\text{i}\rho^+_\text{R}\right),\label{eq:rxx1}\\
    \bar{G}^{(0)}_{1,1}&=&\frac{1}{|t_\text{R}|} \left( \omega^-_\text{R} - \text{i} \rho^-_\text{R}\right),\label{eq:rxx2}
\end{eqnarray}
where we define $\omega_\text{R}^{\pm}=(\omega\pm \epsilon_\text{R})/(2|t_\text{R}|)$ and $\rho_\text{R}^{\pm}={\rm{Re}}\sqrt{1-(\omega_\text{R}^{\pm})^2}$.
The bare Green functions of the left lead $G^{(0)}_{0,0}$ and $\bar{G}^{(0)}_{0,0}$ again have the same structure and one only needs to replace the coupling strengths $t_\text{R}$ and $\epsilon_\text{R}$ of the right lead with their respective counterparts in the left chain. The lesser bare Green functions can again be found by considering $G_{1,1}^{(0)<} = f_\text{R}(\omega) (G_{1,1}^{(0)R}-G_{1,1}^{(0)A})$ and corresponding procedures for the hole Green function and their respective versions of the left lead.  Now Eq.~\eqref{eq:fullgf0} applied to the $\Delta_i=0$ case yields
\begin{eqnarray}
        G_{0,0}&=&\frac{1}{(|t_\text{L}|\omega_\text{L}^+-|t_\text{R}|\omega_\text{R}^+) + \text{i} \left(|t_\text{L}|\rho_\text{L}^++|t_\text{R}|\rho_\text{R}^+\right)},\\
        \bar{G}_{0,0}&=&\frac{1}{(|t_\text{L}|\omega_\text{L}^--|t_\text{R}|\omega_\text{R}^-) + \text{i} \left(|t_\text{L}|\rho_\text{L}^-+|t_\text{R}|\rho_\text{R}^-\right)},
\end{eqnarray}
and when evaluating lesser the Green functions we find
\begin{eqnarray}
        G_{0,0}^<&=& -2\text{i} \frac{ f_\text{L}| t_\text{L}|\rho^+_\text{L}+f_\text{R}|t_\text{R}|\rho^+_\text{R}}{\mathcal{N}^+},\\
        \bar{G}_{0,0}^<&=& -2\text{i} \frac{ f_\text{L}| t_\text{L}|\rho^-_\text{L}+f_\text{R}|t_\text{R}|\rho^-_\text{R}}{\mathcal{N}^-},\\
        \mathcal{N}^{\pm}&=& (|t_\text{L}|\omega_\text{L}^{\pm}-|t_\text{R}|\omega_\text{R}^{\pm})^2+(|t_\text{L}|\rho_\text{L}^{\pm}+|t_\text{R}|\rho_\text{R}^{\pm})^2.
\end{eqnarray}
Having computed all necessary ingredients we are now able to compute the energy current
\begin{eqnarray}
        \dot{E}_\text{L}&=&\frac{t_\text{R}^2}{\pi}\int \text{d} \omega\,\omega\left[f_\text{R}(\omega)-f_\text{L}(\omega)\right] \sum_{\sigma=\pm}\frac{\rho^{\sigma}_\text{R}|t_\text{L}|\rho^{\sigma}_\text{L}}{|t_\text{R}|\mathcal{N}^{\sigma}}\\
                &=&\frac{2}{\pi}\int \text{d}\omega\,\omega\,[f_\text{R}(\omega)-f_\text{L}(\omega)]\frac{|t_\text{L}|\rho_\text{R}^+|t_\text{R}|\rho_\text{L}^+}{\mathcal{N}^+},\label{eq:currentxx}
\end{eqnarray}
where we used that $\omega_\text{L/R}^-(\omega)=-\omega^+_\text{L/R}(-\omega)$. Note that \eqref{eq:currentxx} is valid for arbitrary paramters $t_\text{L/R}$ and $\epsilon_\text{L/R}$ as long as $\Delta_\text{L}=\Delta_\text{R}=0$.

Now specialising to the gapless regime $|\epsilon_\text{L/R}|<2t_\text{L/R}$ and taking the low-temperature limit we obtain
\begin{equation}
 \dot{E}_\text{L}=\frac{\pi}{3}\frac{t_\text{L}t_\text{R}\rho_\text{L}^+(0)\rho_\text{R}^+(0)\,\left(T_\text{R}^2-T_\text{L}^2\right)}{(\epsilon_\text{L}-\epsilon_\text{R})^2/4+(t_\text{L}\rho_\text{L}^+(0)+t_\text{R}\rho_\text{R}^+(0))^2},
 \label{eq:xxtrinv}
\end{equation}
where $\rho_\text{L/R}^+(0)=\sqrt{1-[\epsilon_\text{L/R}/(2t_\text{L/R})]^2}$. The dependence on $t_\text{L}/t_\text{R}$ for $\epsilon_\text{L}=\epsilon_\text{R}=0$ was confirmed by DMRG calculations (not shown).

We note that \eqref{eq:xxtrinv} becomes maximal in the translational invariant case, $t_\text{L}=t_\text{R}$, $\epsilon_\text{L}=\epsilon_\text{R}$, with the result $\dot{E}_\text{L}=\frac{\pi}{12}(T_\text{R}^2-T_\text{L}^2)$. This finding is independent of the value of the on-site potential $\epsilon_\text{L/R}$ and thus the effective velocity at low energies, and it is exactly twice the energy current obtained in the case of coupled Majorana chains \eqref{eq:Isingfinal}. The additional factor of two between Eqs.~\eqref{eq:Isingfinal} and \eqref{eq:xxtrinv} can be traced to the different central charge of the two systems, where here we have $c=1$ instead of the previous $c=1/2$. In particular, in the low-temperature regime the result \eqref{eq:xxtrinv} is again in agreement with the field-theoretical result \eqref{eq:conjecture}.

\subsection{Coupling a fermion chain to a Majorana chain}
The final setup we consider in this work is a system that couples a chain with $\Delta_\text{L}=0$ on the left side to a chain with $\Delta_\text{R}=t_\text{R}$ on the right side. For simplicity, we assume $\epsilon_\text{L}=0$, implying that the left chain is critical with the velocity of the low-energy modes given by $v=2t_\text{L}$. For $\epsilon_\text{R}=2t_\text{R}$ we thus consider a situation in which a critical theory with central charge $c_\text{L}=1$ is coupled to one with central charge $c_\text{R}=1/2$. Thus obviously translational invariance is broken and the result \eqref{eq:conjecture} is not applicable.

The energy current between two systems possessing different central charges has been considered in two previous works. First, Bernard et al.\cite{Bernard-15} directly considered the coupling of two different conformal field theories. Their construction imposes a specific boundary condition on the stress tensor at the boundary between the conformal field theories, which unfortunately cannot simply be related to a condition on the coupling Hamiltonian $H_\text{C}$ appearing in our microscopic setup. A more recent work by Mazza et al.\cite{Mazza-18} used the generalised hydrodynamic approach to study the energy transport in the critical $\mathbb{Z}_3$ parafermionic chain (equivalent to the three-state quantum Potts chain). A peculiarity of this model allowed the study of a setup coupling two critical systems with central charges $c_\text{L}=1$ and $c_\text{R}=4/5$ with the former being held at negative temperature. Based on formal analogies and numerical simulations they conjectured the energy current to behave as $J_\text{E}=J_\text{E}^\infty+\frac{\pi}{12}(c_\text{L}T_\text{L}^2+c_\text{R}T_\text{R}^2)$ with $J_\text{E}^\infty$ being a non-universal contribution and the relative sign originating from the presence of a negative temperature.

Coming back to our setup, we can follow the derivation of the energy current of the coupled Majorana chains up to the point where the explicit results for the left Green functions need to be plugged in. The result for the bare Green function for the left lead is again obtained the same way as Eqs.~\eqref{eq:rxx1} and~\eqref{eq:rxx2} and reads
\begin{equation}\label{eq:L0xx}
    G^{(0)}_{0,0}=\bar{G}^{(0)}_{0,0} \frac{1}{|t_\text{L}|}\left(\omega_\text{L}-\text{i}\rho_\text{L}\right)\equiv\mathcal{L}^{(0)},
\end{equation}
where $\omega_\text{L}=\omega/(2|t_\text{L}|)$ and $\rho_\text{L}={\rm{Re}}\sqrt{1-\omega_\text{L}^2}$ since we have assumed $\epsilon_\text{L}=0$. While the bare Green function of the left chain does not possess anomalous terms, the full Green function acquires off-diagonal terms by virtue of the coupling to the right lead where anomalous terms exist. Solving Eq.~\eqref{eq:fullgf0} and from it constructing $\mathcal{L}$, we find
\begin{equation}\label{eq:Lxx}
    \mathcal{L}=\frac{2}{t_\text{L}^2\mathcal{L}^{(0)*} -2t^2_\text{R}\mathcal{R}^{(0)}}
\end{equation}
and from this result, using Eq.~\eqref{eq:gldef}, we obtain
\begin{equation}
    \mathcal{L}^{<}=-4\text{i}\frac{t_\text{L}\rho_\text{L}f_\text{L}+2t_\text{R}\rho_\text{R}f_\text{R}}{|t_\text{L}^2\mathcal{L}^{(0)*}-2t_\text{R}^2\mathcal{R}^{(0)}|}.
\end{equation}
Thus, the energy current in this setup is given by
\begin{equation}\label{eq:xx-ising-current}
    \dot{E}_\text{L}=\frac{2}{\pi}\int\frac{\text{d}\omega\,\omega\,t_\text{L}t_{\text{R}}\rho_\text{L}\rho_\text{R} [f_\text{R}(\omega)-f_\text{L}(\omega)]}{(t_\text{L} \omega_\text{L}-2t_\text{R} \omega_\text{R})^2 + (t_\text{L}\rho_\text{L}+2t_\text{R}\rho_\text{R})^2}\;.
\end{equation}
We note that this result is still applicable for arbitrary $\epsilon_\text{R}$, but that we have already set $\epsilon_\text{L}=0$ yielding the simplification $\rho_\text{L}=\rho_\text{L}^\pm=\text{Re}\sqrt{1-\left(\omega/(2t_\text{L})\right)^2}$. Furthermore, the integration range is restricted by the densities $\rho_\text{L/R}$ and the $\delta$-function $\omega=0$ can be neglected with the same reasoning as before.

\begin{figure}[t]
\begin{center}
\includegraphics[width=0.45\textwidth]{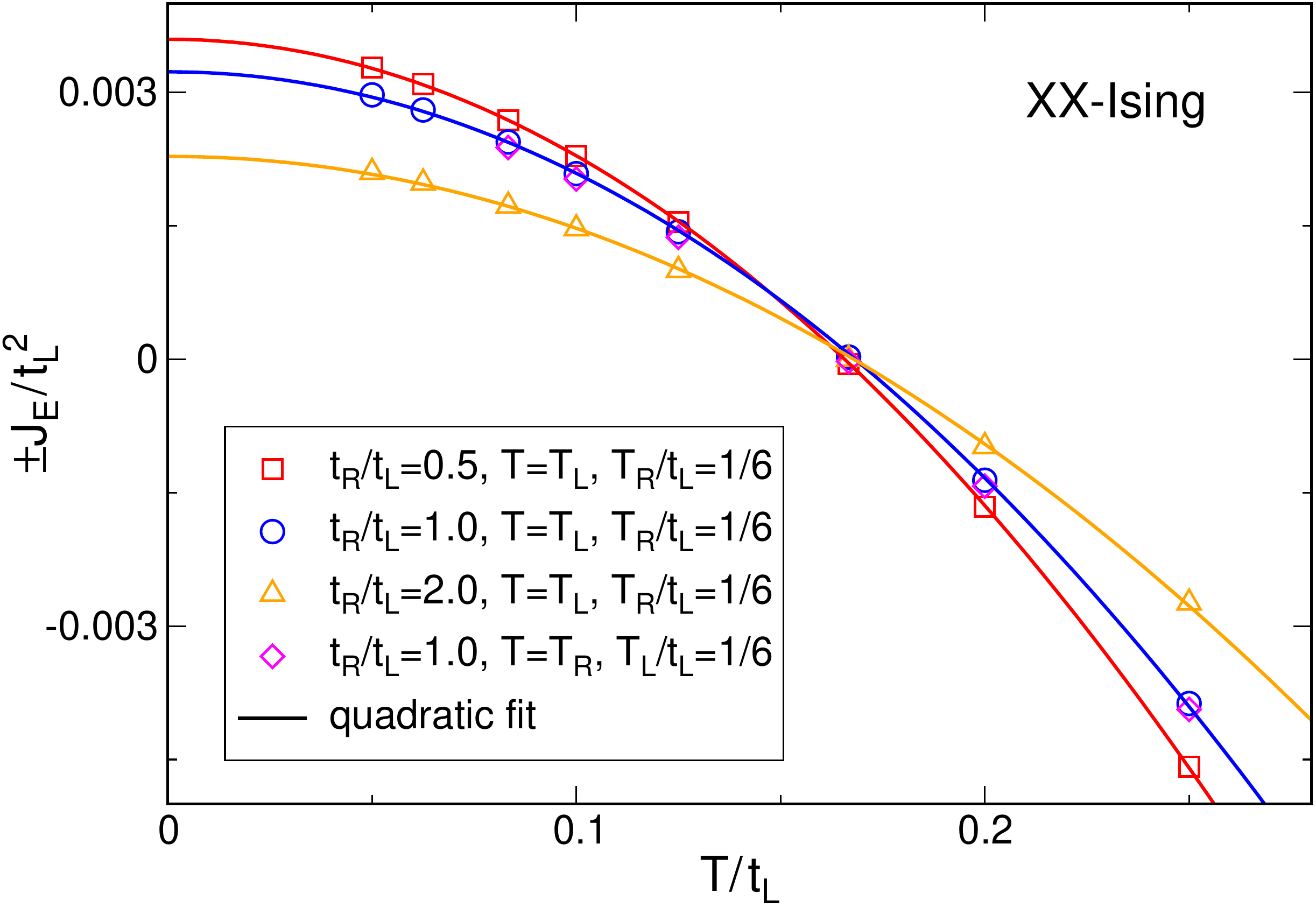}
\caption{Steady-state energy current for the coupling of an XX chain to a critical Ising chain as a function of the temperature in the left or right subsystem. The prefactor in \eqref{eq:XXIsingfinal} is obtained from a quadratic fit, the obtained values are collected in Table~\ref{tab:XXIsing}.}
\label{fig:XXIsing}
\end{center}
\end{figure}

For obvious reasons, this setup does not allow for looking at the special case of translation invariance. However, taking the right chain to be critical, ie, $\epsilon_\text{R}=2t_\text{R}$ allows us to study the case of coupling two critical systems with different central charges. In the low-temperature limit Eq.~\eqref{eq:xx-ising-current} simplifies according to
\begin{equation}
\dot{E}_\text{L}=\frac{\pi}{3}\frac{t_\text{L}/t_\text{R}}{(2+t_\text{L}/t_\text{R})^2}\left(T_\text{R}^2-T_\text{L}^2\right).
\label{eq:XXIsingfinal}
\end{equation}
This becomes maximal for $t_\text{L}=2t_\text{R}$ with the value $\dot{E}_\text{L}=\frac{\pi}{24}(T_\text{R}^2-T_\text{L}^2)$, thus representing the maximal energy current in this setup. This is also confirmed by the numerical data shown in Fig.~\ref{fig:XXIsing} as well as the extracted prefactors in Table~\ref{tab:XXIsing}. We find excellent agreement between the two approaches.

\begin{table}[b]
\caption{Comparison of the analytic and numerical results for the prefactor of the quadratic temperature dependence of the energy current \eqref{eq:XXIsingfinal}.}
\begin{center}
\begin{tabular}{|c|c|c|}
\hline
$t_\text{L}/t_\text{R}$ & analytic value & fitted value\\
\hline
0.5 & 0.084 & 0.081 \\
1 & 0.116 & 0.114 \\
2 & 0.131 & 0.131 \\
\hline
\end{tabular}
\end{center}
\label{tab:XXIsing}
\end{table}

To conclude, we have found the maximal energy current for the free fermion chain coupled to the Majorana chain to be given by \begin{equation}
J_\text{E}=\frac{\pi}{24}(T_\text{L}^2-T_\text{R}^2)=\frac{c\pi}{12}(T_\text{L}^2-T_\text{R}^2)
\label{eq:JEfinalresult}
\end{equation}
with $c=1/2$ being the central charge of the Majorana system.
This can be interpreted such that the transport is limited by the system with the least transport capability, ie, fewer degrees of freedom. Our result, Eq.~\eqref{eq:JEfinalresult}, is consistent with the result obtained by Bernard et al.\cite{Bernard-15} where a U(1) conformal field theory with $c=1$ was coupled to a Majorana field theory ($\mathbb{Z}_2$ parafermion theory).
On the other hand, in contrast to the conjecture put forward by Mazza et al.\cite{Mazza-18} our result depends only on one of the central charges.

\section{Conclusion and Outlook}\label{sec:conclusion}
In this work we studied the energy transport in one-dimensional critical systems which are characterised by differing central charges. The model we studied was a p-wave superconductor which allows to realise two different conformal field theories in its low-energy limit: a theory of free fermions corresponding to central charge $c=1$ and free Majorana fermions corresponding to $c=1/2$. In translationally invariant systems we verified, using an exact non-equilibrium Green function calculation as well as DMRG simulations, that the low-energy limit of the energy current is indeed given by Eq.~\eqref{eq:conjecture}.

Going beyond this we considered systems consisting of two different semi-infinite chains realising a $c=1$ and $c=1/2$ conformal field theory. Since translation invariance is broken in this setup, the result \eqref{eq:conjecture} is not applicable. Nevertheless, our result \eqref{eq:JEfinalresult} for the maximal energy current has the same functional form with the central charge of the Majorana system limiting the energy current. One can interpret our result in the sense that the subsystem possessing the smaller central charge and thus the fewer degrees of freedom limits the energy transport, in the same way that the number of open transport channels limits charge currents in the Landauer--B\"uttiker formalism. In such a picture the limited number of degrees of freedom in the Majorana system thus serves as a bottleneck for the transport through the junction. With this interpretation in mind we conjecture that the maximal energy current between two critical systems possessing central charges $c_\text{L}$ and $c_\text{R}$ should have the form
\begin{equation}
J_\text{E}=\frac{\pi}{12}\,\min(c_\text{L},c_\text{R})\,(T_\text{L}^2-T_\text{R}^2).
\label{eq:ourconjecture}
\end{equation}
For the future it would be interesting to verify that this bottleneck effect also shows up in interacting systems like the XXZ Heisenberg chain and systems corresponding to conformal field theories with central charge $c\neq 1$. An example for the latter setup would be provided by coupling a critical three-state Potts chain to Majorana fermions.

\section{Acknowledgements}

We thank Denis Bernard, Axel Cort\'{e}s Cubero, Benjamin Doyon, Michael Wimmer, Jacopo Viti and particularly Tatjana Pu\v{s}karov for useful discussions. DS would like to thank the organisers of the 2016 programme ``Mathematical aspects of quantum integrable models in and out of equilibrium" at the Isaac Newton Institute for Mathematical Sciences, where this work was partially inspired. This work is part of the D-ITP consortium, a program of the Netherlands Organisation for Scientific Research (NWO) that is funded by the Dutch Ministry of Education, Culture and Science (OCW). CK acknowledges support by the Deutsche Forschungsgemeinschaft through the Emmy Noether program (KA 3360/2-1).

\appendix
\section{Energy currents in the spin chain representation}\label{app:spinchains}
The numerical calculation of the energy current is done in the formulation of the system \eqref{eq:ham} in terms of spin chains, which is obtained by performing the Jordan--Wigner transformation
\begin{eqnarray}
\sigma_j^x&=&\prod_{k<j}\bigl(1-2c_k^\dagger c_k\bigr)\bigl(c_j^\dagger +c_j\bigr),\\
\sigma_j^y&=&-\text{i}\prod_{k<j}\bigl(1-2c_k^\dagger c_k\bigr)\bigl(c_j^\dagger -c_j\bigr),\\
\sigma_j^z&=&2c_j^\dagger c_j-1.
\end{eqnarray}
For example, for the most interesting case of coupling a fermion chain to a critical Majorana chain we obtain with $\Delta_\text{L}=\epsilon_\text{L}=0$, $\Delta_\text{R}=\epsilon_\text{R}/2=t_\text{R}$ and $t=\Delta=t_\text{R}$
\begin{eqnarray}
H_\text{L}&=&-\frac{t_\text{L}}{2}\sum_{j\le -1}\bigl(\sigma_j^x\sigma_{j+1}^x+\sigma_j^y\sigma_{j+1}^y\bigr),\\
H_\text{R}&=&-t_\text{R}\sum_{j\ge 1}\bigl(\sigma_j^x\sigma_{j+1}^x+\sigma_j^z\bigr)+\text{const},\\
H_\text{C}&=&-t_\text{R}\sigma_0^x\sigma_1^x.
\end{eqnarray}
Thus the energy current to be evaluated numerically is given by
\begin{equation}
\dot{E}_\text{L}=\text{i}\Big\langle\bigl[H_\text{C},H_\text{L}\bigr]\Big\rangle=-t_\text{L}t_\text{R}\big\langle\sigma_{-1}^y\sigma_0^z\sigma_1^x\big\rangle.
\end{equation}
Similarly, for the coupling of two critical Majorana/Ising chains we obtain
\begin{equation}
\dot{E}_\text{L}=-2t_\text{L}t_\text{R}\big\langle\sigma_0^y\sigma_1^x\big\rangle.
\end{equation}

\section{The generating functional}\label{sec:genfun}
In this section we will show how to to rewrite the expectation values in Eq.~\eqref{eq:current0} in terms of non-equilibrium Green functions. The non-equilibrium setting of our system requires the use of the Schwinger--Keldysh technique. For brevity of notation we consider a more generic Hamiltonian that encompasses the class of systems Eq.~\eqref{eq:ham} as special cases:
\begin{equation}
    H=\sum_{i,j=-N_L}^{N_R} \left[c^{\dagger}_i T_{i,j} c^{\phantom{\dagger}}_j +c_i^\dagger \frac{\Delta_{i,j}}{2} c^\dagger_j + c_j^{\phantom{\dagger}} \frac{\Delta^{*}_{j,i}}{2} c_i^{\phantom{\dagger}}\right],
\end{equation}
where the factors of 1/2 account for double counting that in this formulation occurs for the off-diagonal terms only.
In order to apply standard field theory techniques it is useful to rewrite this Hamiltonian in a Nambu basis as
\begin{eqnarray}
    H&=& \frac{1}{2}\sum_{i,j=-N_L}^{N_R} \begin{pmatrix} c_i^{\dagger} & c_i \end{pmatrix} \mathcal{H}_{i,j} \begin{pmatrix} c_j \\ c^{\dagger}_j \end{pmatrix}
        +\sum_{j=-N_L}^{N_R} T_{i,i},\qquad\label{eq:hamextended}  \\
    \mathcal{H}_{i,j}&=& \begin{pmatrix} T_{i,j} & \Delta_{i,j} \\ \Delta_{i,j}^* & -  T_{i,j} \end{pmatrix} \, .
\end{eqnarray}
The last term constitutes a global energy shift that has no impact on the physics of our problem, thus we may neglect it. Also note that the prefactor 1/2 is not included in the definition of $\mathcal{H}_{i,j}$

The path integral of this system is formulated on the closed Schwinger--Keldysh contour. In order to derive expressions for the Green functions we introduce generating functionals according to
\begin{equation}
    \mathcal{Z}\left[\bar{\xi},\xi\right]=\int \mathcal{D}\bar{c} \;\mathcal{D} c \;e^{\text{i}S[\bar{c},c,\bar{\xi},\xi]}
\end{equation}
where $S[\bar{c},c,\bar{\xi},\xi]$ is the action to be specified in Eq.~\eqref{eq:action} below. Furthermore, we introduce the fields $c_{i,\alpha}$, where $\alpha=\pm$ specifies the branch of the Schwinger--Keldysh contour on which the field is defined. Here $\alpha=+$ ($\alpha=-$) implies the forward (backward) branch. Introducing the short hand notation $ C_{i}^{\alpha}(t)= \left( c_{i}^{\alpha}(t),\bar{c}_{i}^{\alpha}(t) \right)^{T}$ and $\Xi = \left( \xi_{i}^{\alpha}(t),-\bar{\xi}_{i}^{ \alpha}(t) \right)^T$ as well as adopting the Einstein summation convention, the action reads
\begin{eqnarray}
        & &\text{i}S\left[\bar{C}, C, \bar{\Xi}, \Xi\right] = \frac{\text{i}}{2}\int \text{d}t\,\text{d}t'\,\bar{C}_{i}^{\alpha}(t)\left[ \hat{G}^{-1}\right]_{i,j}^{\alpha,\beta} C_{j}^{\beta} \nonumber\\
         & &\qquad
         + \frac{1}{2}\int\text{d}t\,\text{d}t'\,\left[\bar{\Xi}_{i}^{\alpha}(t) \mathcal{A}_{i,j}^{\alpha,\beta}(t,t')C_{j}^{\beta}(t')\right. \nonumber\\
        & &\qquad\qquad\qquad\qquad\left.+ \bar{C}_{i}^{\alpha}(t) \mathcal{A}_{i,j}^{\alpha,\beta}(t,t')\Xi_{j}^{\beta}(t,t')\right],\qquad\label{eq:action}
\end{eqnarray}
where the inverse Green function is given by
\begin{equation}\label{eq:invgreen}
\left(\hat{G}^{-1}\right)_{i,j}^{\alpha,\beta}= \left[ \text{i} \partial_t \delta_{i,j} \mathds{1} - \mathcal{H}_{i,j}\right]\delta(t-t') \,\tau_z^{\alpha,\beta}
\end{equation}
and
\begin{equation}
    \mathcal{A}_{i,j}^{\alpha,\beta}(t,t')=\delta_{i,j}\tau_z^{\alpha,\beta}\mathds{1}\,\delta(t-t').
\end{equation}
 Here, $\tau_z$ (third Pauli matrix) acts on the contour indices while the unity operator acts in Nambu space.

It is important to note that the inverse Green function Eq.~\eqref{eq:invgreen} appears diagonal in contour subspace $\alpha,\beta$. However, this is an artefact of the continuum time notation\cite{Kamenev11} with the inverse possessing off-diagonal element fixed by imposing appropriate boundary conditions.

Following a standard procedure we integrate out the fermion fields (note that $\mathcal{Z}_0 = 1$ in the non-equilibrium framework because of the closed time contour) leading to
\begin{eqnarray}
        \mathcal{Z} &= & e^{\text{i}S_{\rm{eff}}},\label{eq:seff}\\
        S_{\rm{eff}}&=&-\frac{1}{2}\int\text{d}t\,\text{d}t'\,\bar{\Xi}^{\alpha}_{i}(t)\hat{G}^{\alpha,\beta}_{i,j}(t,t') \Xi^{\beta}_j(t').
\end{eqnarray}
In this expression we introduced the Green function $\hat{G}$, which is not just the inverse of Eq.~\eqref{eq:invgreen} in the sense that it is not diagonal in contour space. In Nambu space, the Green function has the form Eq.~\eqref{eq:nambustructure}. The expectation values needed for the energy current can now be obtained from Eq.~\eqref{eq:seff} by means of appropriate differentiations, ie,
\begin{eqnarray}
        \langle c^{\dagger  +}_i(t) c_j^{-}(t') \rangle &=& \left. \frac{\delta^2\mathcal{Z} }{\delta \xi^{+}_i(t) \delta \bar{\xi}^{-}_{j}(t')}\right|_{\xi=\bar{\xi}=0} \nonumber\\
                    &=&\frac{\text{i}}{2}\left(G^{-+}_{j,i}(t,t') - \bar{G}^{+-}_{i,j}(t',t)\right)\nonumber\\
                    &=&\frac{\text{i}}{2}\left( G^{<}_{j,i}(t,t')+ \left[\bar{G}^{<}_{j,i}(t,t')\right]^*\right),\label{eq:expectationvalues1}\\
        \langle c^{+}_i(t) c^{\dagger -}_{j}(t')\rangle &=& \left. \frac{\delta^2 \mathcal{Z}}{\delta \bar{\xi}^{+}_i(t) \delta \xi^{-}_{j}(t')}\right|_{\xi= \bar{\xi}=0}\nonumber\\
                    & =& \frac{\text{i}}{2}\left(\bar{G}^{<}_{j,i}(t,t')+\left[G^{<}_{j,i}(t,t')\right]^*\right),\label{eq:expectationvalues2}\\
        \langle c^{\dagger +}_i(t) c^{\dagger -}_j(t') \rangle &=& \left.- \frac{\delta^2\mathcal{Z} }{\delta \xi^{+}_i(t) \delta \xi^{-}_{j}(t')}\right|_{\xi=\bar{\xi}=0}\nonumber\\
                    &=&\frac{\text{i}}{2} \left(\bar{F}^{<}_{j,i}(t,t') + \left[F^{<}_{j,i}(t,t')\right]^*\right),\label{eq:expectationvalues3}\\
    \langle c_i^{+}(t) c_j^-(t') \rangle &= &\left.- \frac{\delta^2\mathcal{Z} }{\delta \bar{\xi}^{+}_i(t) \delta \bar{\xi}^{-}_{j}(t')}\right|_{\xi=\bar{\xi}=0}\nonumber\\
                    & = &\frac{\text{i}}{2}\left(F^{<}_{j,i}(t,t')+\left[\bar{F}^{<}_{j,i}(t,t')\right]^*\right),\qquad\label{eq:expectationvalues4}
\end{eqnarray}
where we used the relation $\hat{G}^{<}=\hat{G}^{-+}=-[\hat{G}^{+-}]^{\dagger}$ between lesser and larger Green functions. An important feature of these equations is the relative sign between the definitions of the diagonal and off-diagonal terms caused by the anticommutation relations of fermionic fields.

\section{Decomposition of Green functions}\label{sec:todo}
In real space, the retarded Green function of the full system (and its inverse) can be expressed as a matrix whose indices correspond to lattice sites. In this basis, the inverse of the Green function has only entries on the diagonal and the first off-diagonals on both sides
\begin{equation}\label{eq:invgreen2}
    \hat{G}^{-1}= \begin{pmatrix}
                                  \ddots & \ddots & 0 &\dots \\
                                     \ddots & \left(\hat{G}^{(0)-1}\right)_{i,i} &\hat{t}_{i,i+1}& 0 \\
                                     0 & \left(t_{i,i+1}\right)^{\dagger} & \left(\hat{G}^{(0)-1}\right)_{i+1,i+1}& \ddots\\
                                    \vdots &  0  & \ddots & \ddots
\end{pmatrix},
\end{equation}
where the elements of Eq.~\eqref{eq:invgreen2} each are a matrix in Nambu space of the form
\begin{eqnarray}
    \left(\hat{G}^{(0)-1}\right)_{i,i}&=&\begin{pmatrix}
        \omega +\epsilon_{i} & 0 \\ 0 & \omega - \epsilon_{i}
        \end{pmatrix},\\
    \hat{t}_{i,i+1} &=& \begin{pmatrix} t_i & \Delta_i \\ -\Delta_i & -t_i \end{pmatrix},
\end{eqnarray}
and the values of $t_i$, $\Delta_i$ and $\epsilon_i$ have to be taken at their values corresponding to their respective position in the chain. Evaluating the equation
\begin{equation}\label{eq:identity}
    \sum_k \hat{G}_{i,k} \hat{G}^{-1}_{k,j} = \delta_{ij}
\end{equation}
for the choice $j = i+1$ allows us to rewrite a transition Green function in terms of localised Green functions as
\begin{equation}\label{eq:trf}
     \hat{G}_{i,i+1} = \hat{G}_{i,i}\hat{t}_{i,i+1} \hat{G}_{i+1,i+1}^{(0)}.
\end{equation}
If we combine the relations for $j=i+1$ and $j=i-1$ with the result for $j=i$, we further find
\begin{eqnarray}
    \hat{G}_{i,i} &=&\left[\left(\hat{G}^{(0)-1}\right)_{i,i} - \hat{t}_{i,i+1}\hat{G}^{(0)}_{i+1,i+1} \left(\hat{t}_{i,i+1}\right)^\dagger \right.\nonumber\\*
                   & &\qquad\qquad\left. - \left(\hat{t}_{i-1,i}\right)^{\dagger}\hat{G}^{(0)}_{i-1,i-1}\hat{t}_{i-1,i}\right]^{-1}.\qquad\label{eq:fullgf0}
\end{eqnarray}
Each localised Green function can thus be expressed by localised bare Green functions. Note that this is only possible due to a lack of long-ranged interactions. It is also important to realise that in this context $\hat{G}^{(0)}_{i\pm1,i\pm 1}$ describes the endpoint of a semi-infinite chain that terminates before site $i$ either coming from the left or the right side of the chain.

In order to solve the endpoints of a bare chain, we consider the endpoints of a matrix of the form~\eqref{eq:invgreen2}, multiply with its inverse and solve for either of the appropriate Green functions. For a chain that starts at site $i=1$ and goes on to the right side, we find
\begin{equation}
    \hat{G}_{1,1}^{(0)}=\left[\left(\hat{G}^{(0)-1}\right)_{1,1} - \hat{t}_{1,2} \hat{G}^{(0)}_{2,2}\left(\hat{t}_{1,2}\right)^{\dagger}\right]^{-1},
\end{equation}
where $\hat{G}^{(0)}_{2,2}$ again describes a semi-infinite chain that now starts at site $i=2$. Since a semi-infinite chain that is shortened by one site is still essentially a semi-infinite chain, we may shift $\hat{G}^{(0)}_{2,2}=\hat{G}^{(0)}_{1,1}$ and find the self-consistency equation for semi-infinite chains to the right
\begin{equation}\label{eq:selfconsistencyr}
    \hat{G}^{(0)}_{1,1} =\left[\left(\hat{G}^{(0)-1}\right)_{1,1} - \hat{t}_{1,2} \hat{G}^{(0)}_{1,1}\left(\hat{t}_{1,2}\right)^{\dagger}\right]^{-1}
\end{equation}
and, applying the same procedure to a left-side semi-infinite chain,
\begin{equation}\label{eq:selfconsistencyl}
    \hat{G}^{(0)}_{0,0}= \left[\left(\hat{G}^{(0)-1}\right)_{0,0} - \left(\hat{t}_{-1,0}\right)^{\dagger}\hat{G}^{(0)}_{0,0}t_{-1,0}\right]^{-1}.
\end{equation}


\end{document}